\documentclass[prb,amsmath,amssymb]{revtex4}

\usepackage{graphicx}
\usepackage{dcolumn}
\usepackage{bm}

\newcommand{\mapright}[1]{\smash{\mathop{\hbox to 1.0cm{\rightarrowfill}}\limits^{#1}}}

\begin{document}

\preprint{}

\title{Quantum Information Processing and Entanglement in Solid State Devices}

\author{S. Kawabata\footnote{Electronic address: s-kawabata@aist.go.jp}}
\affiliation{%
Nanotechnology Research Institute and Research Consortium for 
Synthetic Nano-Function Materials Project (SYNAF), National Institute of 
Advanced Industrial Science and Technology (AIST), 1-1-1 Umezono, Tsukuba, 
Ibaraki 305-8568, Japan
}%

\date{\today}

\begin{abstract}
Control over electron-spin states, such as coherent manipulation, filtering and measurement promises access to new technologies in conventional as well as in quantum computation and quantum communication. 
In this paper, we review recent theoretical  proposal of using electron spins in quantum confined structures as qubits.
 We also present a theoretical proposal for testing Bell's inequality in nano-electronics devices.
We show that the entanglement of two electron spins can be detected in the spin filter effect in the nanostructure semiconductor / ferromagnetic semiconductor / semiconductor junction.
In particular, we show how to test Bell's inequality via the measurement of the current-current correlation function in this setup.
\end{abstract}

\maketitle

\section{Introduction}
\label{intro}
Quantum computation (QC)\cite{rf:NC} has attracted much
interest recently, as it opens up the possibility
of outperforming  classical computation through new and more powerful
quantum algorithms such as the ones discovered by Shor\cite{rf:Shor} and
by Grover\cite{rf:Grover}. There is now a growing list of quantum tasks\cite{rf:task}
such as
secret sharing, error correction schemes,  quantum
teleportation, {\it etc.},  that have indicated even more the desirability
of experimental implementations of QC.
In QC the state of
each bit is allowed to be any state of a quantum two-level
system---a qubit, and QC proceeds by one and two-qubit
gates by which all quantum algorithms can be implemented.
There are now a
number of  schemes which have been proposed to realize physical
implementations of qubits and quantum gates.
Here we will review
recent qubit proposal based on the spin of electrons confined in
quantum dots (section 2). 
One of the essential features of such qubits is that they are scalable to many
qubits, that recent experiments demonstrated very long spin  decoherence
times in semiconductors, and that the qubit defined as electron-spin is mobile and thus can be used
for implementing quantum communication schemes.

On the other hand, entanglement, or quantum nonlocality between quantum systems, is  a remarkable feature of quantum mechanics which gives rise to striking phenomena such as the violation of Bell's inequality\cite{rf:Bell1,rf:Bell2}.
Bell's inequality has already been tested experimentally with photons, i.e., massless particles.\cite{rf:Bellexp1,rf:Bellexp2,rf:Bellexp3}
To date, however, no experiments have been reported for massive particles such as electrons.
The semiconductor nanofabrication techniques have allowed us to test the foundations of quantum mechanics in nano-scale solid state devices.
Recent experimental studies include  the fermionic two-particle interferometry (electron antibunching experiment)\cite{rf:tpi}  and the Hanbury Brown and Twiss experiment\cite{rf:HBT1,rf:HBT2} in semiconductor nanostructures.
However, these phenomena are not based on the nature of entangled particles.
Although many methods to generate a spin-entangled electron pair in nano-scale devices have been proposed,\cite{rf:entangle1,rf:entangle2,rf:entangle3,rf:entangle4}  there exists no clear theoretical proposal for the test of the violation of Bell's inequality for spin-entangled electrons in these systems.
In section 3, we shall show that nano-scale semiconductor / ferromagnetic semiconductor / semiconductor (S/FS/S) systems provide a possibility to test Bell's inequality\cite{rf:Kawabata}.  

\section{Spintronics and Quantum Dots for Quantum Computing }
\label{Spin}
In 1998, Loss and DiVincenzo showed a theoretical proposal of electron spin quantum computer (see Fig. 1)\cite{rf:quantumdot1}.
In this proposal, quantum dots play a central role.
The electrodes (dark gray)
 confine single electrons to the dot regions (circles).
The electrons can be moved by electrical gating into the
 magnetized or high-$g$ layer to produce locally different
Zeeman splittings. 
Alternatively, such local Zeeman fields
can be produced by magnetic field gradients as e.g.\ produced
by a current wire.
Since every dot-spin is subject to a different Zeeman splitting,
  the spins can be addressed individually, e.g.\ 
  through ESR pulses of an additional in-plane magnetic AC field
  with the corresponding Larmor frequency.
Such mechanisms can be used for single-spin rotations and the
initialization
step.
The exchange coupling between the dots is controlled by electrically
lowering
 the tunnel barrier between the dots.

%
%
%
\begin{figure}[t]
\begin{center}
\includegraphics[width=10cm]{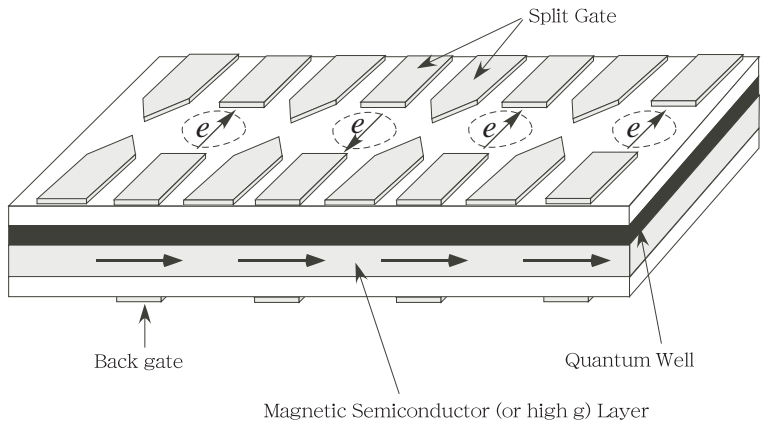}
\end{center}
\caption{Solid state spin quantum computer.}
\label{f1}
\end{figure}
%
%
%
%
Two coupled electrons in zero magnetic field
 have a spin-singlet ground state,
 while the first excited state in the presence of strong Coulomb repulsion
 is a spin triplet.
Higher excited states are separated from these two lowest states
 by an energy gap,
 given either by the Coulomb repulsion or the single-particle confinement.
The low-energy dynamics of such a system can be described by the
 effective Heisenberg Hamiltonian
\begin{equation}\label{Heisenberg}
H_{\rm int}(t)=J(t)\,\,{\bf S}_1\cdot{\bf S}_2,
\end{equation}
where $J(t)$ denotes the exchange coupling between
the two spins ${\bf S}_{1}$ and ${\bf S}_{2}$,
 i.e.\ the energy difference between the triplet and the singlet.
After a pulse of $J(t)$ with
$\int_0^{\tau_s} dtJ(t)/\hbar = J_0\tau_s/\hbar = \pi$ (mod $2\pi$),
the time evolution
$U(t) = T\exp(i\int_0^t H_{\rm int}(\tau)d\tau/\hbar)$
corresponds to the swap operator $U_{\rm sw}$.
While $U_{\rm sw}$ is not sufficient for quantum computation,
 any of its square roots $U_{\rm sw}^{1/2}$,
  turns out to be a universal quantum
 gate.
This is shown by constructing the known universal gate CNOT\cite{rf:CNOT},
 through combination of
$U_{\rm sw}^{1/2}$ and
 single-qubit operations,
 applied in the sequence,
\begin{equation}
U_{\rm CNOT} = e^{-i(\pi/2)S_1^y} e^{i(\pi/2)S_1^z} e^{-i(\pi/2)S_2^z} U_{\rm sw}^{1/2}
e^{i\pi S_1^z} U_{\rm sw}^{1/2} e^{-i(\pi/2)S_1^y}.
\end{equation}
Thus, it can be used, together with single-qubit rotations,
 to assemble any quantum algorithm.

\section{Test Bell's inequality in Semiconductor Nano-Structures}
\label{Bell}

In this section, we shall show that nano-scale semiconductor / ferromagnetic semiconductor / semiconductor (S/FS/S) systems provide a possibility to test Bell's inequality.
It was shown that the spin decoherence time for electrons in semiconductors is very long, i.e., on the order of microseconds\cite{rf:Kikkawa}.
Therefore, the electron spin in these systems becomes a good candidate for investigating Bell's inequality in a solid-state environment.
The scheme proposed by us here consists of an entangler and the S/FS/S junction which act as a spin-polarized beam splitter (SPBS).
We shall show how to generate and detect spin-entangled states experimentally.
As we shall see, this setup allows for a direct test of Bell's inequality.
For this purpose, we calculate the current-current correlation function using the quantum scattering theory and compare it with the result of a local hidden variable (LHV) theory\cite{rf:Bell2,rf:LHV}.
We also discuss the effect of imperfection of the SPBS in order to make a clear comparison with experiments in the future.

\begin{figure}[t]
\begin{center}
\includegraphics[width=6cm]{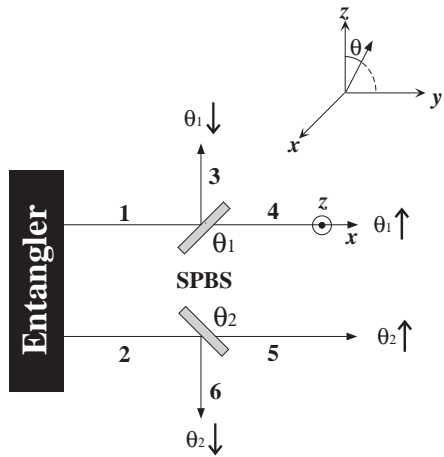}
\end{center}
\caption{Schematic setup for a test of the violation of Bell's inequality in nano-sclale spintronics devices.
Spin-entangled electrons generated from the entangler are fed into the spin-polarized beam splitters (SPBSs) with the polarization angle $\theta_{1(2)}$ through the lead wires $\alpha=1,2$.
The entanglement of a spin singlet can be detected in a current correlation measurement at the output leads ($\alpha=3,4,5,6)$.}
\label{f2}
\end{figure}
%
In the following, we propose a setup which involves the entangler and two S/FS/S junctions, see Fig. 2.
The entangler is assumed to be a device that produces pairs of electrons in an entangled spin singlet\cite{rf:entangler}, the specific realization being the coupled-semiconductor quantum dots, each of which contains a 
single electron spin\cite{rf:quantumdot1,rf:quantumdot2,rf:quantumdot3,rf:quantumdot4,rf:quantumdot5}.
The key element of this proposal is the SPBS which ensures the spin-up (down) electron leaving the entangler to be transmitted (reflected), i.e., the spin filter effect\cite{rf:spinfilter1,rf:spinfilter2}.
It was shown that ferromagnetic semiconductors, in particular EuS\cite{rf:EuS1,rf:EuS2} and EuSe\cite{rf:EuSe}, can be grown in thin films which exhibit the strong spin filter effect.
In Fig. 3, we show schematic energy band diagrams for the S/FS/S junction.
Because of the exchange splitting of the electron barrier in the FS layer, spin-up electrons will tunnel through the barrier easily while spin-down electrons will not.
In favorable cases of EuSe, the spin polarization $
P
\equiv
(T_{\uparrow}-T_{\downarrow})/(T_{\uparrow}+T_{\downarrow})
$
($T_{\uparrow(\downarrow)}$ is the transmission probability for spin-up (down) electrons)  in tunneling has exceeded $97\%$\cite{rf:EuSe}.
In order to test Bell's inequality, we must change the relative polarization direction $\theta_2-\theta_1$ between the two FS layers arbitrarily.
This can be easily achieved by magnetizing two layers along different directions before measurement.
\begin{figure}[t]
\begin{center}
\includegraphics[width=6cm]{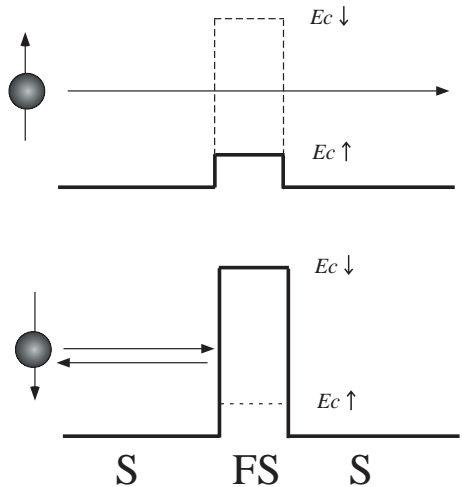}
\end{center}
\caption{Conduction band profile of the spin filter semiconductor / ferromagnetic semiconductor / semiconductor (S/FS/S) junction.
$E_{c\uparrow(\downarrow)}$ is the conduction band energy for spin up (down) electrons in the FS layer.
Exchange interaction gives rise to a spin-dependent potential; spin down electrons see a large barrier while spin up electron a small one.
}
\end{figure}

By extending the standard quantum scattering theory\cite{rf:entangler,rf:scattering1,rf:scattering2}, we shall calculate  the current-current correlation function for the entangled incident spin-entangled state.
We assume that the incident state is given by the spin singlet,
\begin{eqnarray}
\left|  \psi \right> 
=
\frac{1}{\sqrt{2}} 
\left[  
            a^{\dagger}_{1 \uparrow} \left( E_1 \right)  a^{\dagger}_{2 \downarrow} \left( E_2 \right)
			-
			 a^{\dagger}_{1 \downarrow} \left( E_1 \right)  a^{\dagger}_{2 \uparrow} \left( E_2 \right)
\right]
\left|  0 \right>
,
  \label{eqn:e1}
\end{eqnarray}
where $\left|  0 \right>$ denotes the filled Fermi sea, i.e., the electron ground state in the leads and $a^{\dagger}_{\alpha \sigma} \left( E \right)$ creates an incoming electron in the input leads $\alpha(=1,2)$ with spin $\sigma$ (along the $z$ direction) and energy $E$.
Note that since we deal with discrete energy states here, we normalize the operator $a_{\alpha \sigma} \left( E \right)$ such that 
\begin{eqnarray}
\left[ 
           a_{\alpha \sigma} \left( E \right)
		   ,
		   a^{\dagger}_{\beta \sigma'} \left( E' \right)
\right]
=
\delta_{\sigma,\sigma'} \delta_{\alpha,\beta} \delta_{E,E'}/\nu
,
  \label{eqn:e2}
\end{eqnarray}
where $\nu$ is the density of states in the lead wires.
We assume that each lead wire consists of a single quantum channel.
Thus, the current operator in the output leads $\alpha(=3,4,5,6)$ is given by
\begin{eqnarray}
I_{\alpha \sigma}  \left( \theta, t \right)
=
\frac{e}{h\nu} \sum_{E,E'}
\left[
          a^{\dagger} _{\alpha \sigma} \left(\theta,E \right)
          a _{\alpha \sigma} \left(\theta,E' \right)
-          b^{\dagger} _{\alpha \sigma} \left(\theta,E \right)
          b _{\alpha \sigma} \left( \theta,E' \right)
\right]
\exp \left[  i \frac{E-E'}{\hbar} t\right]
.
  \label{eqn:e3}
\end{eqnarray}
The operator $a^{\dagger} _{\alpha \sigma} \left(\theta,E \right)$ creates the electron with spin $\sigma$ along a specified direction with the polar angle $\theta$ (see Fig. 2). 
Then, the relationship between operator $a _{\alpha \sigma} \left(\theta,E \right)$ and $a _{\alpha \sigma} \left(E \right)$ is given by the spinor transformation
\begin{eqnarray}
\left(
\begin{array}{c}
     a _{\alpha \uparrow} \left(\theta ,E\right) \\
     a _{\alpha \downarrow} \left(\theta  ,E\right)  
\end{array}
\right)
=
\left(
\begin{array}{cc}
 \cos   \frac{\theta}{2}  & \sin  \frac{\theta}{2}  \\
-\sin   \frac{\theta}{2}  & \cos   \frac{\theta}{2} 
\end{array}
\right)
\left(
\begin{array}{c}
      a _{\alpha \uparrow}  \left(E\right)\\
     a _{\alpha \downarrow}\left(E\right)
\end{array}
\right)
.
  \label{eqn:e4}
\end{eqnarray}
In eq. (\ref{eqn:e3}), the operator $b_{\alpha \sigma} \left(\theta,E \right)$ is related to the operator $a_{\alpha \sigma} \left(\theta,E \right)$ via the scattering matrix $s_{\alpha\sigma,\beta\sigma}$,
\begin{eqnarray}
b _{\alpha \sigma} \left(\theta,E \right)
 =\sum_{\beta=1}^6 s_{\alpha\sigma,\beta\sigma}
  a _{\beta \sigma} \left(\theta,E \right)
  .
  \label{eqn:e5}
\end{eqnarray}
%
%
%
Using eqs.~(\ref{eqn:e3}) and ~(\ref{eqn:e5}), we arrive at the following expression for the current operator
\begin{eqnarray}
I_{\alpha \sigma} \left( \theta, t \right)
&=&
\frac{e}{h\nu} 
\sum_{E,E'}
\sum_{\beta,\gamma=1}^6
a^{\dagger} _{\beta \sigma} \left(\theta,E \right)
A^{\alpha}_{\beta\gamma} \left( \sigma \right)
a_{\gamma \sigma} \left(\theta,E' \right)
\exp \left[  i \frac{E-E'}{\hbar} t\right]
,
  \label{eqn:e6}
\\
A^{\alpha}_{\beta\gamma} \left( \sigma \right)
&\equiv&
\delta_{\alpha,\beta}\delta_{\alpha,\gamma}
-
s_{\alpha\sigma,\beta\sigma}^{*}s_{\alpha\sigma,\gamma\sigma}
.
  \label{eqn:e7}
\end{eqnarray}
The current-current correlation function between the leads $\alpha(=3,4)$ and $\beta(=5,6)$ is given by
\begin{eqnarray}
C_{\alpha\sigma,\beta\sigma'} (\theta_1,\theta_2)
&\equiv&
\equiv
\lim_{t' \rightarrow \infty}
\frac{1}{t'}
\int_{0}^{t'}
d t
\left< \psi \right|   
                I_{\alpha\sigma} \left( \theta_1,t\right)  
				I_{\beta\sigma'}  \left( \theta_2,t\right) 
\left|   \psi  \right>
\nonumber\\
&=&
\frac{e^2}{h^2\nu^2}
\sum_{E,E'}
\sum_{\gamma,\delta,\varepsilon,\xi=1}^6
A^{\alpha}_{\gamma\delta} \left( \sigma \right)
A^{\beta}_{\varepsilon\xi} \left( \sigma' \right)
\left< \psi \right|   
                a^{\dagger} _{\gamma \sigma} \left(\theta_1,E \right)
				a _{\delta \sigma} \left(\theta_1,E \right)
                a^{\dagger} _{\varepsilon \sigma'} \left(\theta_2,E' \right)
				a _{\xi \sigma'} \left(\theta_2,E' \right)
\left|   \psi  \right>
.
  \label{eqn:e8}
\end{eqnarray}
Substituting $\left| \psi \right>$ defined in eq.~(\ref{eqn:e1}) into eq.~(\ref{eqn:e8}) and using the commutation relation eq.~(\ref{eqn:e2}), we obtain
\begin{eqnarray}
C_{4\uparrow,5\uparrow} (\theta_1,\theta_2)
=
C_{3\downarrow,6\downarrow} (\theta_1,\theta_2)
=
\frac{e^2}{2h^2\nu^2} 
\sin^2 \left( \frac{\theta_1-\theta_2}{2} \right)
,
  \label{eqn:e9}
\\
C_{4\uparrow,6\downarrow} (\theta_1,\theta_2)
=
C_{3\downarrow,5\uparrow} (\theta_1,\theta_2)
=
\frac{e^2}{2h^2\nu^2} 
\cos^2 \left( \frac{\theta_1-\theta_2}{2} \right)
.
  \label{eqn:e10}
\end{eqnarray}
Therefore, by measuring these correlation functions, we can detect entanglement between electron-spins in two lead wires.
In  order to compare the above quantum mechanical calculation with the LHV theory\cite{rf:LHV}, we consider the following function;
\begin{eqnarray}
F \left( \theta_1,\theta_2\right)
\equiv
\frac{
           \displaystyle{
           C_{4\uparrow,5\uparrow} (\theta_1,\theta_2)
		   +
		   C_{3\downarrow,6\downarrow} (\theta_1,\theta_2)
		   -
		   C_{4\uparrow,6\downarrow} (\theta_1,\theta_2)
		   -
		   C_{3\downarrow,5\uparrow} (\theta_1,\theta_2)
          }
		  }
		  {
          \displaystyle{
           C_{4\uparrow,5\uparrow} (\theta_1,\theta_2)
		   +
		   C_{3\downarrow,6\downarrow} (\theta_1,\theta_2)
		   +
		   C_{4\uparrow,6\downarrow} (\theta_1,\theta_2)
		   +
		   C_{3\downarrow,5\uparrow} (\theta_1,\theta_2)
		  }
		  }
		  .
		  \nonumber\\
  \label{eqn:e11}
\end{eqnarray}
Assuming that the two SPBS are ideal (
$
T_\uparrow
\equiv
\left| s_{4\uparrow,1\uparrow}\right|^2
=
\left| s_{5\uparrow,2\uparrow}\right|^2=1
$
, 
$
R_\downarrow
\equiv
\left| s_{3\downarrow,1\downarrow}\right|^2
=
\left| s_{6\downarrow,2\downarrow}\right|^2=1
),  
$
we  obtain 
\begin{eqnarray}
F \left( \theta_1,\theta_2\right)
		  =
		  -\cos \left( \frac{\theta_1-\theta_2}{2} \right)
		  .
  \label{eqn:e12}
\end{eqnarray}
Bell showed that any LHV theory must obey the inequality,
\begin{eqnarray}
\left|     B \left( \theta_1,\theta_2,\theta'_1,\theta'_2\right)   \right|
\equiv
\left|  
           F \left( \theta_1,\theta_2\right) 
		- F \left( \theta_1,\theta'_2\right)  
		+ F \left( \theta'_1,\theta_2\right)
		+ F \left( \theta'_1,\theta'_2\right)    
\right|
\le 2
.
\nonumber\\
   \label{eqn:e13}
\end{eqnarray}
This inequality is called Bell's inequality.
On the other hand, our quantum mechanical calculation gives the result
\begin{eqnarray}
B \left( \theta_1,\theta_2,\theta'_1,\theta'_2\right)
=
-\cos \left( \frac{\theta_1-\theta_2}{2} \right)
+\cos \left( \frac{\theta_1-\theta'_2}{2} \right)
-\cos \left( \frac{\theta'_1-\theta'_2}{2} \right)
-\cos \left( \frac{\theta'_1-\theta_2}{2} \right)
.
  \label{eqn:e14}
\end{eqnarray}
By choosing, $\theta_1-\theta_2=\theta'_1-\theta'_2=\theta'_1-\theta_2=(\theta_1-\theta'_2)/3\equiv\Theta$, one obtains 
\begin{eqnarray}
B \left( \Theta \right)
=
\cos 3 \Theta -3 \cos \Theta
.
  \label{eqn:e15}
\end{eqnarray}
When $\Theta=3\pi/4$, $B=2\sqrt{2}$ shows clear violation of Bell's inequality $\left| B \right| \le2$ (see Fig. 4).
%
%
%
\begin{figure}[b]
\begin{center}
\includegraphics[width=6cm]{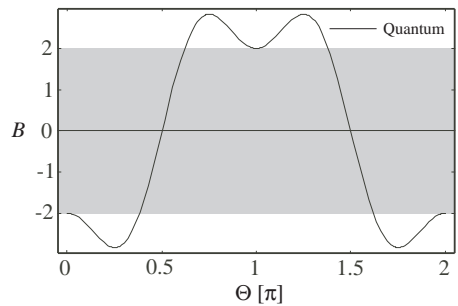}
\end{center}
\caption{The quantum correlation (solid line) as a function of the relative angle $\Theta$ of the spin-polarized beam splitters and the region which satisfy the Bell's inequality (shaded area).}
\end{figure}
%
%
%
%

In the above calculation, we have assumed that the SPBS is ideal, i.e., the transmission probability $T_{\uparrow}$ for spin-up electrons equals one, and that the reflection probability $R_{\downarrow}$ for spin-down electrons also equals one.
However, in practice it is difficult to fabricate a perfect SPBS, because the electron affinity of the S layer is generally different from that of the FS layer\cite{rf:EuS1}.
This gives rise to imperfect transmission for spin-up electrons, i.e., $T_{\uparrow}<1$.
Moreover, as a result of a finite barrier height for spin-down electrons in real junctions,  $R_{\downarrow}$ becomes less than unity.
Therefore, it is important to determine how small the values $T_{\uparrow}$ and $R_{\downarrow}$ can be to still have the violation of  Bell's inequality, i.e., $\left| B \right|>2$.
Below, we shall derive the condition for $T_{\uparrow}$ and $R_{\downarrow}$ by extending the above calculation.

In the case of an imperfect SPBS, the current operator in each output lead is expressed as the sum of contributions from spin-up and spin-down electrons, i.e,  $I_{\alpha}(\theta,t) = \sum_{\sigma=\uparrow,\downarrow}I_{\alpha,\sigma}(\theta,t)$.
By calculating the current-current correlation functions and using eq.~(\ref{eqn:e11}), we obtain 
\begin{eqnarray}
B (\Theta)
=
\frac{1}{2}
\left[
             \left\{
			                 \left( T_{\uparrow} - R_{\uparrow}  \right)^2 
							 +
			                 \left( T_{\downarrow} - R_{\downarrow}\right)^2 
			 \right\}
\left(
          3 \sin^2  \frac{\Theta}{2} - \sin^2 \frac{3\Theta}{2}
\right)
\right.
+
\left.
			 2
			                 \left( T_{\uparrow} - R_{\uparrow}  \right)
			                 \left( T_{\downarrow} - R_{\downarrow}\right)
\left(
          3 \cos^2  \frac{\Theta}{2} - \cos^2 \frac{3\Theta}{2}
\right)
\right]
.
  \label{eqn:e20}
\end{eqnarray}
Therefore, to have the violation of Bell's inequality, i.e., $B>2$ at $\Theta=3\pi/4$, $T_{\uparrow}$ and $R_{\downarrow}$ must satisfy the following condition:
\begin{eqnarray}
\frac{c}{2}
\left[
(2 T_{\uparrow} -1)^2
+
(2 R_{\downarrow} -1)^2
\right]
+
\frac{c-2}{2}
(2 T_{\uparrow} -1)
(2 R_{\downarrow} -1)
>
1
,
  \label{eqn:e21}
\end{eqnarray}
where $c \equiv 3 \sin^2  (3\pi/8) - \sin^2 (9\pi/8)$ and we have used $T_{\downarrow}=1-R_{\downarrow}$ and $R_{\uparrow}=1-T_{\uparrow}$.
Figure 5 shows the region which is forbidden by the Bell's inequality, i.e., eq.~(\ref{eqn:e13}).
Therefore, we must use the S/FS/S junctions which satisfy the condition of eq.~(\ref{eqn:e21}) to examine experimentally the violation of Bell's inequality.
Particularly, in the case of $T_\uparrow=R_\downarrow$, the condition for $B>2$ is given by
\begin{eqnarray}
T_\uparrow=R_\downarrow
>
\frac{1}{2} + \frac{1}{2\sqrt{c-1}}
\approx
0.92
.
  \label{eqn:e22}
\end{eqnarray}
This corresponds to the  spin polarization $P >84\%$.
Such a spin filter effect can be realized with current spin electronics technology\cite{rf:EuSe}.
%
%
%
%
%
%
\begin{figure}[t]
\begin{center}
\includegraphics[width=6cm]{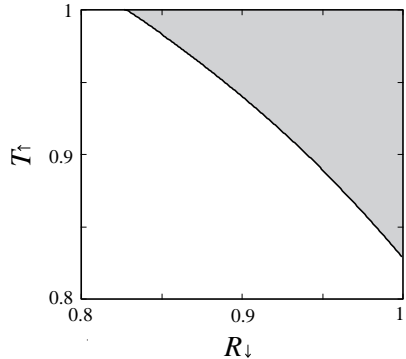}
\end{center}
\caption{Diagram of the condition for $T_{\uparrow}$ and $R_{\downarrow}$ (eq.~(\ref{eqn:e21})) in the case of $\Theta=3\pi/4$.
The shaded region is forbidden by the Bell's inequality.}
\end{figure}
%
%
%

%
%
%
%
\section{Summary}
We have described a concept for a quantum computer
 based on electron spins in quantum-confined nanostructures, in particular
quantum dots,
 and presented theoretical proposals  for testing Bell's inequality in such structures.

By measuring the current-current correlation in the nanostructure system described here, one can probe the entanglement of electron spins.
This can be used as an experimental test of the violation of Bell's inequality.
Spin-entangled electrons produced by the entangler, i.e., coupled quantum dots, enter the SPBS, i.e., the spin filter S/FS/S junctions.
We have calculated the function $B$ for this setup using the quantum mechanical scattering theory and compared it with the result of the LHV theory.
Moreover, for comparison with future experiments,we have also derived the condition for testing the violations of Bell's inequality in the case of an imperfect SPBS.

The device setup described here will be of immediate use in further experimental tests on the foundations of quantum mechanics (e.g., Greenberger-Horne-Zeilinger correlation\cite{rf:GHZ1,rf:GHZ2}, quantum teleportation\cite{rf:teleportation} and quantum eraser experiment\cite{rf:eraser}) and on the quantum information technology (e.g., quantum computation\cite{rf:q-comp} and quantum cryptography\cite{rf:crypto}) in nano-scale solid state devices.

I would like to thank S. Abe, H. Akinaga and  S. Yamada for useful discussion.
This work was supported by NEDO under the Nanotechnology Materials Program.

\end{document}